\def \deg          {\ifmmode ^{\circ}\else $^\circ$\fi}  
\def \etal         {{\it et~al.} }
\def \kms          {\rm{\hbox{km s$^{-1}$}}}
\def \lam          {$\lambda$}

\def \Lya          {\hbox{Ly$\alpha$}}

\def \Msun         {\rm{\hbox{M$_{\odot}$}}}           
\def \mum          {\hbox{$\mu$m}}

\def \pcc          {\hbox{cm$^{-3}$}}

\def \zaz          {{$z_a\kern -1.5pt \approx\kern -1.5pt z_e$}}
\def \zllz         {{$z_a\kern -3pt \ll\kern -3pt z_e$}}
\def \Zsun         {\rm{\hbox{Z$_{\odot}$}}}           
\documentstyle[11pt,paspconf]{article}

\begin{document}
\title{Probing the High-Redshift Universe with Quasar Elemental 
Abundances}
\author{Fred Hamann}
\affil{Center for Astrophysics \& Space Sciences, University of California 
-- San Diego, La Jolla, CA 92093-0424}

\begin{abstract}
The heavy elements near QSOs provide unique measures of the 
star formation and chemical evolution in young galactic 
nuclei or pre-galactic condensations. Studies 
of quasar abundances also naturally address a 
variety of questions regarding the physics of QSO environments, 
including the Baldwin Effect emphasized at this meeting. Here I 
review the status of quasar abundance work.
\end{abstract}

\section{Introduction} 
Quasars (or QSOs) at all redshifts have strong emission and (sometimes) 
absorption lines due to metals in their immediate environments. 
The gas must have undergone some amount of chemical 
enrichment, presumably via local star formation. 
At the highest quasar redshifts, approaching $z\sim5$ 
(Schneider, Schmidt \& Gunn 1991), the enrichment time scales 
cannot be long; the Universe itself was less than 
$(1+z)^{-1}\approx 17$\% of its present age at $z=5$, or roughly 
1 billion years old (depending on the cosmology, see Fig. 1). 
Quasar metal abundances can therefore provide valuable constraints 
on the properties of star formation in the early Universe, 
possibly for the first stars forming in young galactic 
nuclei or pre-galactic condensations. These constraints will be  
important complements to other studies of high-redshift galaxies, 
involving, for example, the ``Lyman-break'' objects (Steidel 
\etal 1998, Connolly \etal 1997) or 
damped-\Lya\ absorbers (Pettini \etal 1997, 
Lu, Sargent \& Barlow 1998), that probe more extended structures 
or rely on very different data and analysis techniques. An 
important goal of quasar research is therefore 
to merge the QSO abundance 
results with these other studies to develop a more complete picture 
of star formation and galaxy evolution at early cosmological epochs. 

Quasar abundance studies also naturally 
address a variety of problems concerning the QSOs 
themselves, such as the circumstances of QSO formation 
and evolution, the location, geometry, dynamics and physical conditions 
of the emission and absorption line regions, the 
relationships between the various emission and absorption phenomena 
(including the soft X-ray ``warm'' absorbers), and the influence of 
metallicity on the physics and observable properties of QSO 
environments. The last of these items specifically involves the 
Baldwin Effect (see Osmer \& Shields and Korista \etal in this 
volume). 

Three independent probes of QSO abundances are 
readily observable at all redshifts: the broad emission lines (BELs), 
the broad absorption lines (BALs) and the intrinsic narrow 
absorption lines (NALs). Each diagnostic has its own 
theoretical and observational uncertainties, so it is 
important to consider as many of them as possible. Here 
I review the status of QSO abundance studies based on these 
diagnostics. I will emphasize my own ongoing 
work with various collaborators. 
Please see Hamann \& Ferland (1999 -- hereafter HF99) for a more 
thorough review of this topic. 

\begin{figure}[h]
\plotfiddle{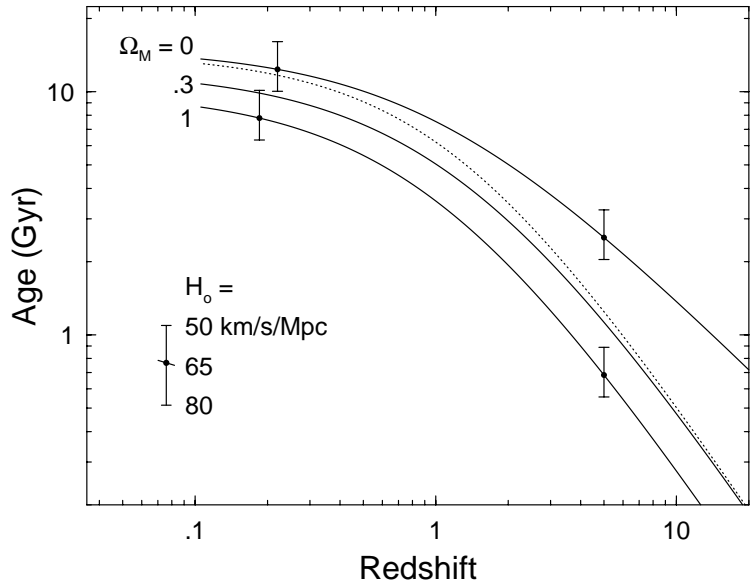}{2.8in}{0.0}{45.0}{45.0}{-180.0}{-380.0}
\caption{Redshift versus age of the Universe 
in Big Bang cosmologies. The three solid curves correspond to 
$H_o = 65$~km~s$^{-1}$~Mpc$^{-1}$, $\Omega_{\Lambda} = 0$ and 
$\Omega_{M} = 0$, 0.3 and 1. The dotted curve corresponds 
to $\Omega_{\Lambda} = 0.7$ and $\Omega_{M} = 0.3$. 
The ``error'' bars show the range of ages possible for $H_o$ between 
50 and 80~km~s$^{-1}$~Mpc$^{-1}$ (see Carroll \& Press 1992).}
\end{figure}

\section{Quasars and Galaxy Evolution}

QSOs are believed to reside in the cores of massive 
galaxies or, perhaps at the highest redshifts, in dense 
condensations that later become the cores of massive galaxies. 
See, for example, Kormendy 
\etal (1998), Magorrian \etal (1998) for the black hole--host galaxy 
connection, McLure \etal (1998), McLeod (1998), 
Aretxaga, Terlevich \& Boyle (1998), 
Bahcall \etal (1997), Miller, Tran \& Sheinis (1996), 
McLeod \& Rieke (1995) for direct observations 
of QSO hosts, and Turner (1991), Haehnelt \& Rees (1993), 
Loeb \& Rasio (1994), Haehnelt, Natarajan 
\& Rees (1998), Haiman \& Loeb (1998) for theoretical 
links between QSOs and galaxy formation. 

\subsection{Expectations for QSO Abundances}

The intrinsic emission and absorption lines of QSOs 
offer direct probes of the composition and enrichment 
history of the gas in these dense galactic environments. 
Studies of nearby galaxies indicate that vigorous star 
formation in galactic cores should have produced super-solar gas-phase 
metallicities ($Z$) within a few billion years of the initial collapse 
(e.g. Arimoto \& Yoshii 1987; K\"oppen \& Arimoto 1990). 
The enriched gas might ultimately be ejected from the 
galaxy/QSO environment, consumed by the central black hole, or diluted 
by subsequent gaseous infall, but the evidence for early-epoch 
high-$Z$ gas remains in the old stars today.  
In particular, the mean stellar metallicities\footnote{Note that  
``metallicity'' is measured best from the enrichment products 
of massive stars like O, Mg, etc., rather than Fe (see Wheeler 
\etal 1989 and \S2.3 below).} in the cores of 
nearby massive galaxies (including the bulge of our own Milky 
Way) are typically $\sim$1 to 3~\Zsun\ (e.g. Rich 1988, 
Bica, Arimoto \& Alloin 1988, Bica, Alloin \& Schmidt 1990, 
Gorgas, Efstathiou \& Arag\' on Salamanca 
1990, Worthey, Faber \& Jes\' us Gonzalez 1992, 
McWilliam \& Rich 1994, Minniti \etal 
1995, Idiart \etal 1996, Bruzual \etal 1997). Individual stars 
are distributed about these means with metallicities reflecting the 
gas-phase abundance at the time of their formation. Only the most 
recently formed stars at any epoch have metallicities as high as 
that in the gas. Simple chemical evolution models of spheroidal 
systems like elliptical galaxies or the bulges spiral disks, 
wherein the gas-phase metallicity grows monotonically with time 
(Searle \& Zinn 1978, Tinsley 1980, Rich 1990), predict that the 
gas was $\sim$2 to 3 times more metal-rich than the current stellar 
means, or $\sim$2 to 9~\Zsun\ near the end of the main 
star-forming epoch. 
We might therefore expect QSO metallicities up to $\sim$9~\Zsun\ 
as long as most of the local star formation (and enrichment) occurs 
before the QSOs ``turn on'' or become observable. 

\subsection{The Galactic Mass-Metallicity Relation}

Studies of nearby galaxies reveal that higher metallicities 
occur in more massive systems (Faber 1973, Bender, Burstein \& Faber 
1993, Zaritsky, Kennicutt \& Huchra 1994). Massive galaxies 
reach higher $Z$'s because they have deeper gravitational potentials 
and are better able to retain their gas against the building thermal 
pressures from supernovae (Larson 1974). 
Low-mass galaxies eject their gas 
before high $Z$'s are attained. Quasar abundance estimates should  
therefore constrain the gravitational binding energy 
of the local star-forming regions and, 
perhaps, the total masses of the host galaxies. 

\subsection{Fe/$\alpha$ as a Cosmological Clock}

Another relevant result from galactic chemical evolution is 
that the ratio of Fe to $\alpha$-element (O, Ne, Mg, ...) 
abundances can constrain the ages of star-forming systems 
(Wheeler, Sneden \& Truran 1989). 
The Fe/$\alpha$ age constraint follows from the different 
enrichment timescales; $\alpha$ elements come from the 
supernova explosions of short-lived massive stars (primarily 
Type II -- SN II's), 
while Fe has a large contribution from the longer-lived 
intermediate-mass binaries that produce Type Ia supernovae (SN Ia's). 
The subsequent delay in the Fe enrichment, roughly 1 Gyr, does 
not depend on the 
star formation rate or other uncertain parameters of the chemical 
evolution; it depends only on the lifetimes of the SN Ia precursors 
(see \S3 below, also Yoshii, Tsujimoto \& Nomoto 1996). 
The ratio of Fe/$\alpha$ abundances can therefore serve as 
an absolute ``clock'' for constraining the epoch of 
first star formation and, perhaps, the cosmology 
itself (see Hamann \& Ferland 1993a -- hereafter HF93a). 
For example, measurements of high Fe/$\alpha$ (near solar 
or higher) in $z>4$ QSOs would place the epoch of star formation 
beyond the limits of current direct observations ($z>6$) 
and might be problematic for cosmologies with $\Omega_M = 1$ 
(Fig. 1). 

\subsection{Nitrogen Abundances}

A final result from galactic abundance studies is that nitrogen 
can also be selectively enhanced at moderate to high metallicities 
due to ``secondary'' CNO nucleosynthesis. There is growing 
evidence for a substantial ``primary'' N contribution at low $Z$ 
in some objects, based on a plateau in [N/O] at roughly $-$0.7 
for [O/H]~$\la -0.7$ in galactic HII regions\footnote{I 
use the notation for logarithmic abundances relative to solar, 
$[a/b] = \log(a/b) - \log(b/a)_{\odot}$}. 
But at higher metallicities 
(in the regime relevant to QSOs, see below), secondary production 
dominates and [N/O] grows roughly in proportion to [O/H] 
(HF93a, Vila-Costas \& Edmunds 1993, Van Zee \etal 1998, 
Izotov \& Thuan 1999). Shields 
(1976) noted that this special behavior should make nitrogen a 
particularly valuable probe of the chemical evolution in QSOs. 

\section{Specific Predictions for Galactic Chemical Evolution}

Chemical enrichment models make specific predictions for the 
gas-phase abundances that can be compared to the QSO data. 
Hamann \& Ferland 1992 and HF93a constructed one-zone models for 
stellar populations assembled by the infall of primordial gas. 
The enrichment follows standard stellar yields that 
compare well with observations of the Milky Way and nearby galaxies.  
The star formation is regulated by power-law initial mass functions 
(IMFs) of 
the form $\Phi\propto M^{-x}$, where $M$ is the stellar mass and 
$\int\Phi$ dM = 1. The enrichment delays caused by finite stellar 
lifetimes are included. We tested the calculations by constructing 
a simple yet viable model of the Galactic solar neighborhood, 
and then varied just the slope of the IMF and the timescales 
for star formation and infall to model the chemical history 
of QSO environments. 

Figure 2 shows the predicted relative abundances 
for two cases at opposite extremes. The ``Solar Neighborhood'' 
model uses a 3~Gyr timescale for the infall of primordial gas and 
an IMF with slope $x=1.6$ for $M$ $\geq$ 1 \Msun\ and  
1.1 for $M$ $<$ 1 \Msun\ (after Scalo 1990). The stellar birth rate 
is set so that $Z$ = 1 \Zsun\ at the time of the sun's formation 
and the fraction of mass in gas is $\sim$15\% at the present epoch. 
The ``Giant Elliptical'' model uses a stellar birth-rate $\ga$100 
times faster and an infall timescale of only 0.05 Gyr 
so that the 
mass fraction in gas is $\sim$15\% after just 0.5 Gyr. 
The IMF is also flatter, with slope $x=1.1$ for all masses. 
The shorter timescales and flatter IMF (more high-mass stars) in 
the Giant Elliptical case produces a rapid evolution to high $Z$'s, 
reaching $\sim$10 \Zsun\ at $\sim$1 Gyr. 
The star formation stops at $\sim$1 Gyr because the 
gas is essentially exhausted; thereafter the system 
evolves ``passively'' and the ejecta from low-mass stars affect 
the abundances somewhat. See HF93a for details. 

The parameters used in these calculations were based on 
standard one-zone infall models of the Galactic disk and 
massive ellipticals (Arimoto \& Yoshii 1987, 
Matteucci \& Tornamb\'e 1987, Matteucci \& Brocato 1990, 
K\"oppen \& Arimoto 1990). However, the results 
are only illustrative and more sophisticated models would 
be needed to match entire galaxies. 

Both models in Figure 2 exhibit the delayed rise and 
subsequent overabundances in N (due to 
secondary CNO processing in stellar envelopes) and Fe (due to 
the delayed enrichment by Type Ia supernova). The late 
increase in Fe/$\alpha$ should be at least 
a factor of a few. The increase is larger in the Giant Elliptical case 
because, by the time SN~Ia's make their Fe contribution, there is 
little gas left in the systems and each SN has a greater 
affect. 

\begin{figure}[h]
\plotfiddle{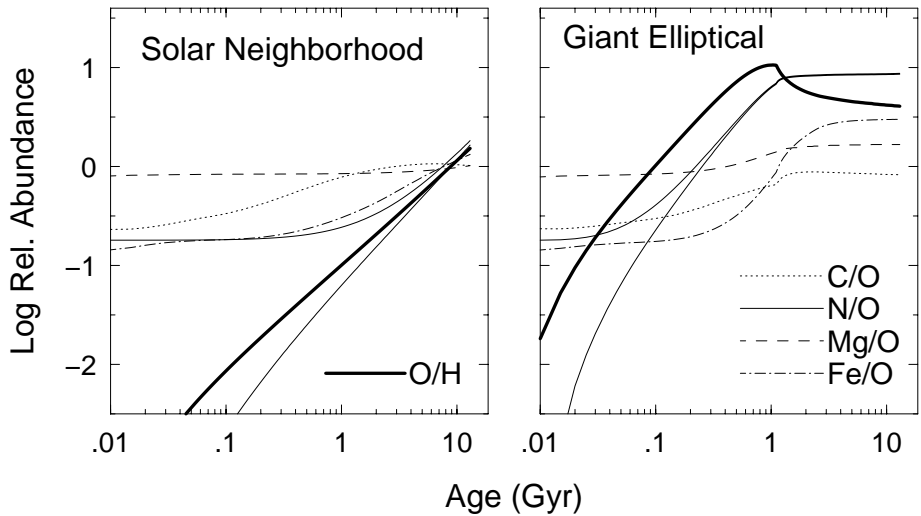}{2.4in}{0.0}{50.0}{50.0}{-200.0}{-465.0}
\caption{Logarithmic 
abundance ratios normalized 
to solar for the two evolution models 
discussed in the text. Two scenarios for the N enrichment 
are shown; one with secondary only and the other with 
secondary+primary (with a plateau in N/O at low $Z$ that is 
forced to fit the HII region data; see \S2.4).}
\end{figure}

\section{Broad Emission Line Diagnostics and Results}

Broad emission lines provide the most readily available 
abundance diagnostics because they are present in all QSOs. 
Many BELs can be measured in large QSO samples using medium-resolution 
spectra. However, the strengths of the metal lines, such as CIV~\lam 1549, 
relative to \Lya\ are surprisingly insensitive to the global 
metallicity (HF99, Hamann \& Ferland 1993b). QSOs could actually 
have a very wide range of 
metallicities while displaying qualitatively similar BEL spectra 
(also Korista \etal 1998). Nonetheless, some line ratios are 
sensitive to the relative abundances. 
The challenge is to identify observable lines that have 
significant abundance sensitivities above their unavoidable 
dependences on other factors. In general, we must rely 
computational models to quantify the various parameter 
sensitivities. Since the BEL region (BELR) spans a range of radii 
(Peterson 1993) and probably has a wide range of densities 
and ionizations (Ferland \etal 1992, 
Baldwin \etal 1995, Hamann \etal 1998), the most important 
criterion for abundance-sensitive line ratios is that they form 
under similar conditions (e.g. in the same or overlapping regions). 
Line ratios involving nitrogen should be particularly useful for 
constraining the total metallicity and evolutionary state of QSO 
environments (\S2.4). Boyle (1990), 
Laor \etal (1995) and Osmer \& Shields (this volume) present 
labeled plots showing the UV emission lines that are generally 
available for abundance studies. 

\subsection{Inter-combination Lines}

Shields (1976, also Baldwin \& Netzer 1978, 
Davidson \& Netzer 1979, Osmer 1980, Uomoto 1984) used several 
ratios involving UV inter-combination (semi-forbidden) lines, 
such as NIII]~\lam 1750/CIII]~\lam 1909, NIII]/OIII]~\lam 1664 and \hfill\\
NIV]~\lam 1486/CIV, to study the relative nitrogen abundances. 
The lines are often weak and difficult to measure, but 
these studies generally concluded that N/O and N/C are 
roughly solar or higher -- consistent with solar or higher 
overall metallicities. That work fell out of 
favor by the mid-1980s as it became clear that BELR densities 
probably exceed the critical densities of the inter-combination 
lines ($\sim$$3\times 10^9$ to $\sim$$10^{11}$~\pcc ). Thus 
collisional deexcitation, which was not previously accounted 
for, can be important. K.T Korista, G.J. Ferland and I have 
just begun a theoretical study to reexamine the usefulness 
of these ratios for abundance work. We suspect that at 
least some of the ratios (e.g. NIII]/OIII]) will prove to be 
useful abundance measures because the lines have similar critical 
densities and the collisional effects should roughly cancel out. 

\subsection{Ratios Involving NV~\lam 1240}

Some early studies of the permitted BELs noted that 
NV~\lam 1240 was significantly stronger than the 
predictions of photoionization models (Osmer \& Smith 
1976, 1977). They noted that this NV enhancement might be 
due to a nitrogen enhancement as discussed by Shields (1976, 
\S4.1). More recently, HF93a, Ferland \etal (1996) and 
Hamann \etal (1997a) performed extensive analysis 
of the NV \lam 1240 emission compared to CIV \lam 1549 
and HeII~\lam 1640 for estimating N/He and N/C abundances. 
The NV/CIV line ratio is advantageous in low signal-to-noise 
spectra because CIV is essentially always present. The 
disadvantage of this ratio is that the NV and CIV line-forming 
regions are not quite coincident. The 
NV/HeII ratio can be harder to measure because HeII is weak, but 
upper limits on HeII still provide useful lower limits on the N/He 
abundance. NV/HeII is a more robust abundance indicator 
because the NV emitting region lies within the He$^{++}$ zone, where 
HeII~\lam 1640 forms by recombination. If N$^{+4}$ does not fill the 
He$^{++}$ zone the NV emission can be weak, but it is not possible to 
produce NV without also producing HeII (see also HF99 for more 
discussion). 

We studied the theoretical NV/HeII and NV/CIV ratios for a wide 
variety of ionizing continuum shapes and BELR parameters. 
We used parameters that maximize (or nearly maximize) these 
ratios for comparisons to the data, 
so that we are most likely to underestimate 
the N/He and N/C abundances and thus the overall metallicity. 
Figure 3 compares the theoretical predictions to line ratios measured 
in QSOs at different redshifts. The predictions   
use abundances from Figure 2,  
a BELR density of 10$^{10}$ \pcc , an incident flux of 
hydrogen-ionizing photons of 10$^{20}$ cm$^{-2}$ s$^{-1}$, and 
the QSO continuum derived by Mathews and Ferland (1987; 
which we altered slightly to have 
$\alpha_{ox} = -1.24$ and an additional decline at 
wavelengths $\ga$1 \mum ).   
This continuum shape produces large but not 
quite maximum NV line ratios. See HF93a, Ferland \etal (1996) 
and Hamann \etal (1997a) for details on the calculations and the 
data set. The evolutionary ages from Figure 2 are converted to 
redshifts assuming the evolution 
begins at the Big Bang in a cosmology with $\Omega_M$ = 1, 
$\Omega_{\Lambda}=0$ and $H_o$ = 65 \kms\ Mpc$^{-1}$.
The results in Figure 3 show that the short timescales, 
flatter IMF (favoring high-mass stars) and higher $Z$'s in the Giant 
Elliptical model provide a much better fit to the high-redshift 
data. Steeper IMFs (more like the Solar 
Neighborhood case) could account for some of the smaller 
line ratios if the evolution times are short enough. 
The largest line ratios at high redshifts 
could be fit by invoking BELR parameters that better optimize 
the NV emission or by using still-flatter IMFs.

\begin{figure}[h]
\plotfiddle{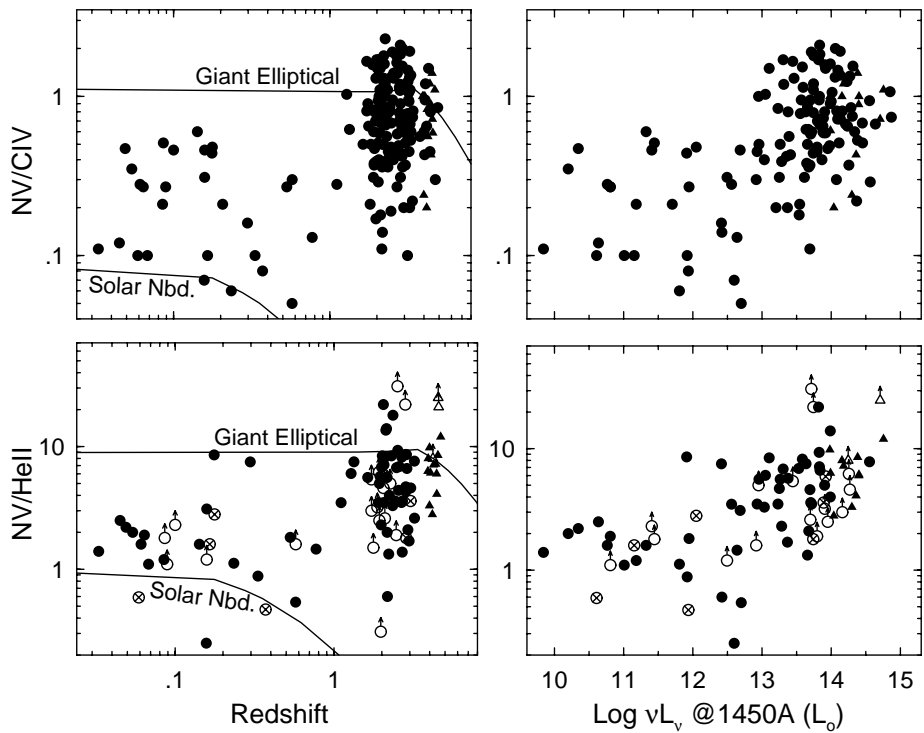}{3.5in}{0.0}{52.0}{52.0}{-212.0}{-440.0}
\caption{Observed and predicted NV/CIV and NV/HeII 
line ratios versus redshift (left panels) and luminosity 
(right panels) for a cosmology with $\Omega_M$ = 1, 
$\Omega_{\Lambda}=0$ and $H_o$ = 65 \kms\ Mpc$^{-1}$ 
(see HF99).}
\end{figure}

One important empirical result is that the NV line ratios 
are typically larger in more luminous QSOs (HF93a, Osmer \etal 1994, 
Laor \etal 1995, V\'eron-Cetty, V\'eron, \& Tarenghi 1983). 
This trend could be affected by relationships between the luminosity 
$L$ and various physical parameters of the BELR; 
however, the trend in the NV line ratios could also 
result entirely from higher metallicities (and higher relative N 
abundances) in more luminous sources (also Korista \etal 1998 
and this volume).  
If QSO luminosities correlate positively with the masses of the 
QSOs and/or their host galaxies (McLeod \& Rieke 1995, 
Magorrian \etal 1998), a metallicity-$L$ 
relationship would indicate that there is a mass-metallicity relation 
among QSOs that is similar 
(or identical) to the well known relation in nearby galaxies (\S2.2). 
The fact that high $Z$'s occur only at high redshifts (Fig. 3) 
might result from the natural tendency to form denser 
or more massive systems at early epochs, when the mean density of 
the Universe was itself higher (Haehnelt \& Rees 1993).  

\subsection{Tests and Uncertainties}

It should soon be possible to test the results based on 
NV for some QSOs using either the intercombination lines or 
other permitted line ratios such as NIII~\lam 990/CIII~\lam 977 
(Hamann, Korista \& Ferland 1999). However, the main conclusion 
for $Z\ga$~\Zsun\ inferred from NV appears to be robust. 
It is not, for example, sensitive to uncertain 
parameters used in the photoionization 
calculations. In fact, adopting other (less than optimal) parameters 
only strengthens the case for large $Z$'s (Ferland \etal 1996). 
Photoionized gas is simply too cool to produce the large observed 
NV/HeII ratios with solar or lower N/He abundances. 
We therefore considered non-radiatively heated clouds, where 
higher temperatures might enhance the collisionally-excited NV line 
relative to the HeII recombination 
transition. We found that collisionally ionized clouds 
can indeed match 
the high NV/HeII ratios with roughly solar abundances 
if the temperatures are above 10$^5$~K. However, this situation 
still underpredicts the NV/CIV ratio -- because both lines are 
collisionally excited -- and overpredicts several other lines (e.g. 
SiIV \lam 1397 and OIV] \lam 1402). Such clouds 
are also thermally unstable and would tend to runaway to coronal 
temperatures (where no NV would be present) on thermal timescales 
of order tens of seconds (Hamann \etal 1995a, Ferland \etal 1996). 
Finally, we also explored the 
possibility that the NV emission is enhanced by scattering in an 
outflowing BAL region rather than by abundance effects (Krolik \& 
Voit 1998, Turnshek \etal 1996, Turnshek 1988). 
Explicit calculations, 
which use observed BAL profiles for input and include scattering 
of both the 
underlying continuum and \Lya\ line radiation, indicate that the 
flux scattered by NV in BAL winds should contribute negligibly 
to the measured NV emission lines (Hamann \& Korista 1996, 
Hamann \etal 1999, HF99). 

\subsection{FeII/MgII and Cosmology}

The observed ratios of MgII~\lam 2800 to FeII (UV) emission 
lines might provide the Fe/$\alpha$ age discriminator discussed 
in \S2.3. The FeII emission spectrum is rich and blended, 
and the emission characteristics are not well understood 
because the Fe$^+$ atom is complex. Nonetheless, 
it is worth investigating because of the potential importance 
of Fe/$\alpha$ abundances. FeII/MgII appears to be the 
best available emission-line diagnostic of Fe/$\alpha$ 
because these lines have similar wavelengths and they should 
form in similar regions. They are also 
measurable in significant numbers of QSOs. 
Previous work on FeII (Wills, Netzer \& Wills 1985) 
suggested that the Fe/Mg abundance ratio is typically
several times larger than solar in QSOs (at moderate redshifts), 
indicating that the star formation began at least $\sim$1 Gyr 
prior to the look-back times of those objects. More recent 
observations indicate that the FeII/MgII line ratios are 
large, consistent with large Fe/Mg abundances, even in the 
highest redshift QSOs (at $z>4$, Thompson, Elston \& Hill 
1999, Taniguchi \etal 1997). New theoretical calculations 
(Verner \etal 1999, Sigut \& Pradhan 1998) of 
the FeII emission from BELRs will further test and quantify  
the FeII/MgII emission lines for abundance purposes.  

\section{Absorption Line Diagnostics and Results}

\subsection{Types of Absorption Lines}

Intrinsic absorption lines provide independent probes of the 
elemental abundances near QSOs that can test and extend the 
emission-line results. Intrinsic absorbers include the BALs, 
at least some of the ``associated'' NALs (with similar absorption 
and emission redshifts, i.e. \zaz ), and any other 
systems that form in (or were ejected from) the vicinity 
of the QSO engine. Figures 4 and 5 show 
examples of QSO spectra containing BALs and NALs, respectively.
\begin{figure}[h]
\plotfiddle{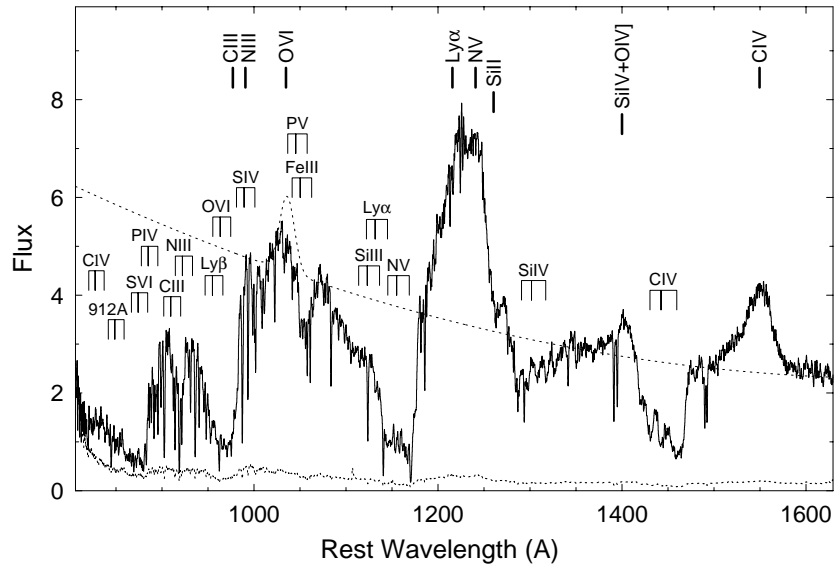}{2.7in}{0.0}{65.0}{65.0}{-200.0}{-190.0}
\caption{Spectrum of the BALQSO PG~1254+047 
($z_e = 1.01$) with 
emission lines labeled across the top and possible BALs marked 
at redshifts corresponding to the 3 deepest minima in  
CIV. Not all of the labeled BALs are present. The smooth dotted 
curve is a continuum fit extrapolated to short wavelengths 
(from Hamann 1998).}
\end{figure}
BALs, with their broad troughs and maximum 
velocity extents often exceeding 10,000~\kms , clearly form in 
high-velocity outflows from the QSOs. NALs  
might form in a variety of environments, ranging from  
QSO winds like the BALs to cosmologically 
intervening gas like galactic halos. 
Each narrow-line system must be examined individually.
Several indicators of intrinsic absorption have been developed, 
including (1) time-variable line 
strengths, (2) line multiplet ratios that imply partial line-of-sight 
coverage of the background light source(s), (3) high space densities 
inferred from excited-state fine-structure lines, and (4) line profiles 
that are broad and smooth compared to thermal line widths 
(see Hamann \etal 1997b, Barlow \& 
Sargent 1997 and references therein). 

\begin{figure}[h]
\plotfiddle{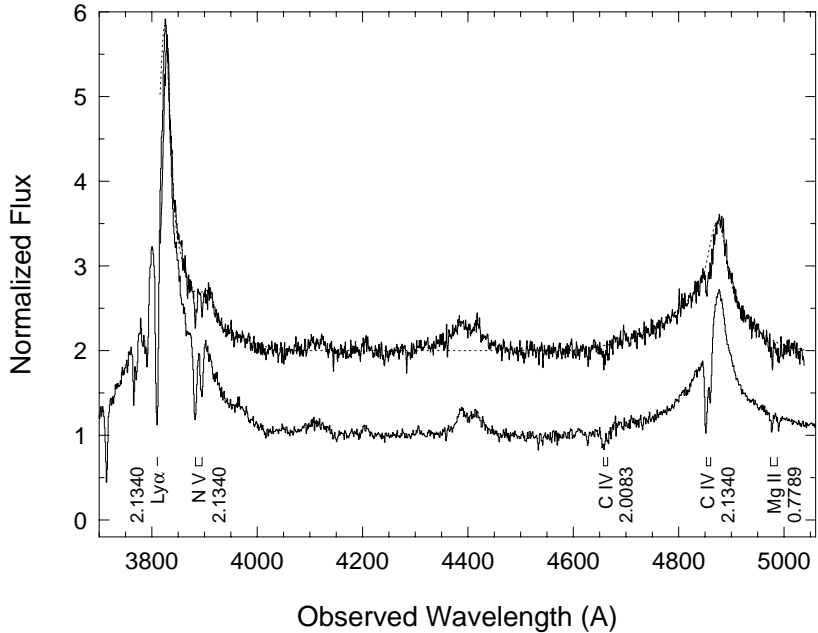}{3.0in}{0.0}{50.0}{50.0}{-205.0}{-420.0}
\caption{Normalized spectra of the QSO UM~675 
($z_e \approx 2.15$) show that its \zaz\ absorption lines 
(at $z_a = 2.134$) 
varied between two observations. The time variability, together 
with the relatively broad line profiles and partial line-of-sight 
coverage revealed by other higher-resolution data, imply that this 
NAL system is intrinsic to the QSO.}
\end{figure}

One well-studied example of an intrinsic narrow-line system is in the 
QSO, UM~675 (Hamann \etal 1995b, 1997b). Figure 5 shows that this 
system varied between two observations. At 
higher spectral resolution ($\sim$9~\kms ) the line profiles appear 
much broader than the thermal speeds (with full widths 
at half minimum of $\sim$470~\kms ) and 
the resolved CIV and NV line troughs appear 
too shallow for the optical depths required by their doublet 
ratios. The troughs are evidently filled-in by 
unabsorbed flux. This filling-in probably results 
from partial coverage of the background light source(s) 
(see HF99 for a sketch of possible 
partial-coverage geometries). The coverage fractions in UM~675 
are $\sim$50\% for CIV and NV and $>$85\% for HI. The variability 
time scale implies 
that the absorber is not more than 1 kpc from the central 
continuum source, and very likely much nearer. 
The diverse absorption lines detected in UM~675 (from CIII~\lam 977 
and NIII~\lam 990 to OVI~\lam 1034 and NeVIII~\lam 774) imply 
a range of ionization states, consistent with a factor of 
$\ga$100 range in density or $\ga$10 in distance from the ionizing 
continuum source (see Hamann \etal 1997b). 

\subsection{General Techniques}

The abundance analysis for absorption lines is, in principle, 
much simpler than for the emission lines because the absorption  
yields direct estimates of the ionic column densities. 
One only has to apply an ionization correction to convert the column 
density ratios into relative abundances. The logarithmic abundance
ratio of any two elements $a$ and $b$ is related to their column 
densities by, 
\begin{equation}
\left[{a\over b}\right] \ = \ \ 
\log\left({{N(a_i)}\over{N(b_j)}}\right) \ +\
\log\left({{f(b_j)}\over{f(a_i)}}\right) \ +\
\log\left({b\over a}\right)_{\odot} 
\end{equation}
where $(b/a)_{\odot}$ is the solar abundance ratio, and $N$ and $f$ 
are  respectively the column densities and ionization fractions 
of elements $a$ and $b$ in ion states $i$ and $j$. 
If the gas is in photoionization equilibrium and optically thin at all 
far-UV continuum wavelengths, the correction factors, $f(b_j)/f(a_i)$, 
depend only on the shape of the ionizing spectrum and the ionization 
parameter $U$ (defined as the dimensionless ratio of the gas to 
hydrogen-ionizing photon densities at the illuminated face of the cloud). 

\begin{figure}[h]
\plotfiddle{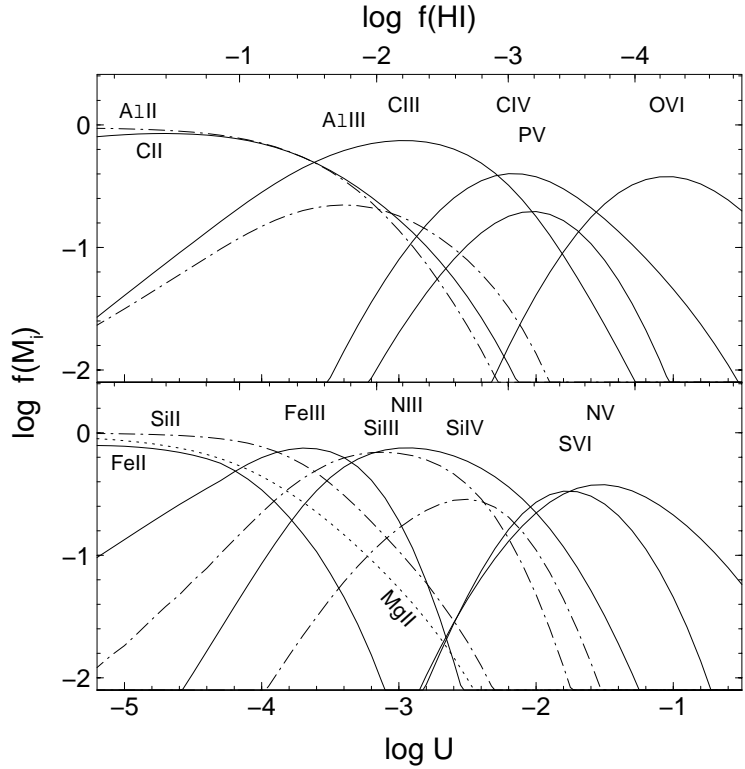}{3.6in}{0.0}{75.0}{75.0}{-240.0}{-225.0}
\caption{Ionization fractions 
in optically thin clouds photoionized at different $U$ by a 
power-law spectrum with index $\alpha =$~$-$1.5. The HI fraction 
appears across the top. The curves for the metal ions $M_i$ 
are labeled at their peaks whenever possible. The notation 
here is SiII = Si$^{+2}$, etc.}
\end{figure}

Figure 6 shows theoretical ionization fractions, $f$(M$_i)$, for 
various metals, M, in ion stage, $i$, as a function of the ionization 
parameter, $U$, in optically thin photoionized clouds (from HF99). 
The HI fraction, $f$(HI), is shown across the top of 
the figure. The calculations were performed using CLOUDY (version 
90.04, Ferland \etal 1998) with a power-law ionizing spectrum 
with index $\alpha = -1.5$, where $f_{\nu}\propto \nu^{\alpha}$. 
Note that the results in Figure 6 are not sensitive to the specific 
densities or abundances used in the calculations (within reasonable 
limits, see Hamann 1997). Ideally, we would constrain the ionization 
state (i.e. $U$) in Figure 6 by comparing the column densities in 
different ionization stages of the same element. We can also constrain the 
ionization by comparing ions of different metals with some reasonable 
assumption about their relative abundance. With $U$ thus constrained, 
Figure 6 provides the correction factors needed 
to derive abundance ratios from Eqn. 1. Repeating this procedure 
with calculations for different ionizing spectral shapes 
yields estimates of the theoretical uncertainties (Hamann 1997). 

If the data provide no 
useful constraints on $U$ (because too few lines are measured) 
or multiple constraints imply a range of ionization states  
(as in the \zaz\ system of UM~675, Hamann \etal 1995b), we can 
still derive conservatively low values of $f$(HI)/$f$(M$_i$) 
and thus [M/H] by assuming each 
metal line forms where that ion is most abundant (i.e. at 
the peak of its $f$(M$_i$) curve in Figure 6). 
We can also place firm lower limits on the [M/H] ratios by 
adopting the minimum correction factor for each M$_i$. 
(Every $f$(HI)/$f$(M$_i$) ratio has a well-defined 
minimum at $U$ values near the peak in 
the $f$(M$_i$) curve.) Hamann (1997) presented numerous 
plots of the minimum and conservatively small ionization 
corrections for a wide range of ionizing spectral shapes. 
Note that the correction factors for some important 
metal-to-metal ion ratios, such as 
FeII/MgII and PV/CIV, also have well-defined minima 
that are useful for abundance constraints (see also 
HF99 and Hamann \etal 1999). 

\subsection{Abundance Results}

Table 1 shows the results of applying this analysis  
to two bona fide intrinsic absorbers, 
namely, the \zaz\ NALs in UM~675 (Hamann 1997; Hamann 
\etal 1997b) and the BALs in PG~1254+047 (Hamann 1998). 
The ionic column densities listed for both sources follow from 
direct integration of the apparent optical depths in the 
absorption line troughs (see \S5.4 below). 
For UM~675, I adjusted the 
optical depths for a 50\% coverage fraction in all of the metal 
ions and 100\% coverage in HI. These adjustments do not change 
the derived HI column density, but the columns in the metal ions are 
roughly doubled. Note 
that the relatively low column densities in both systems (and the lack 
of a significant HI Lyman edge in UM~675) support the assumption of low 
continuum optical depths in the calculations above (Hamann 1997). 

The ionization states in both cases are uncertain, so 
the table lists the conservatively low metal-to-hydrogen 
ratios, [M/H]$_p$, assuming 
each metal line forms at the peak in its $f$(M$_i$) curve. 
The table also lists the firm lower limits, [M/H]$_{min}$, 
derived from the minimum correction factors in Hamann (1997). 
The uncertainties in the table indicate the range of 
values derived for a reasonable range of ionizing continuum shapes. 
Keep in mind that these M/H estimates are limiting values that 
need not agree between ions. 
The [M/H] quantities are typically higher (and more 
realistic) for the low-ionization metals because the HI lines 
tend to form with these ions. The [M/H]$_p$ results provide our 
best guess at the actual abundances in UM~675;  
the overall metallicity is roughly twice  
solar ($Z\approx 2$~\Zsun ) based on [C/H], with nitrogen 
several times more enhanced. The result for $Z\ga$~\Zsun\ 
is typical of \zaz\ NALs and occurs without exception in the intrinsic 
systems (Petitjean, Rauch \& Carswell 1994, Wampler \etal 1993, 
Tripp \etal 1996 and 1997, Savage \& Tripp 1998, 
Hamann 1997, Hamann \etal 1997c). 
The NALs thus support the independent evidence 
from BELs for $Z\ga$~\Zsun\ and enhanced nitrogen (\S4 above).

\begin{table}
\caption{Example Column Densities and Abundances} 
\scriptsize
\vskip 10pt
\begin{tabular}{lcccc}
Ion& log $N$(M$_i$)& & [M/H]$_p$& [M/H]$_{min}$\\
\noalign{\vskip 3pt}
\tableline
\noalign{\vskip 4pt}
\noalign{UM 675 ($z_a\approx z_e\approx 2.15$)}
\noalign{\vskip 3pt}
\tableline
\noalign{\vskip 4pt}
H~I& 14.8&     & ---& ---\cr
\noalign{\vskip 2pt}
C~III& 14.1&     & $+$0.4$^{+0.5}_{-0.2}$ & $-$0.2$^{+0.7}_{-0.2}$\cr
C~IV& 14.9&    & $+$0.4$^{+0.7}_{-0.3}$ & $-$0.1$^{+0.9}_{-0.3}$\cr
N~III& 14.5&    & $+$1.3$^{+0.5}_{-0.2}$ & $+$0.7$^{+0.7}_{-0.3}$\cr
N~V& 15.1&    & $+$0.3$^{+0.8}_{-0.4}$ & $-$0.2$^{+0.9}_{-0.3}$\cr
O~VI& 15.5&    & $-$0.9$^{+0.8}_{-0.3}$ & $-$1.4$^{+0.8}_{-0.3}$\cr
Ne~VIII& 15.7&  & $-$1.1$^{+0.7}_{-0.4}$ & $-$1.4$^{+0.7}_{-0.4}$\cr
\noalign{\vskip 4pt}
\tableline
\noalign{\vskip 4pt}
\noalign{PG 1254+047 (BAL, $z_e = 1.0$)}
\noalign{\vskip 3pt}
\tableline
\noalign{\vskip 4pt}
H~I& \llap{$\la$}15.0&         & ---& ---\cr
C~IV& 15.9& & $+1.4^{+0.7}_{-0.4}$& $+1.0^{+0.8}_{-0.4}$\cr
N~V& 16.2& & $+1.4^{+0.8}_{-0.4}$& $+1.0^{+0.8}_{-0.4}$\cr
Si~IV& 15.0& & $+2.0^{+0.6}_{-0.4}$& $+1.8^{+0.6}_{-0.4}$\cr
P~V& 15.0& & $+3.3^{+0.6}_{-0.4}$& $+3.1^{+0.6}_{-0.4}$\cr
\noalign{\vskip 4pt}
\tableline
\end{tabular}
\end{table}

The derived BAL column densities lead to much more extreme 
abundances, with [Si/H]~$\ga$~1.8 and [P/C]~$\ga$~+2.2 in 
PG~1254+047. These results are roughly typical of BALs 
(Turnshek 1988, Turnshek \etal 1996, Korista \etal 1996, 
Junkkarinen \etal 1997, Hamann 1997). They are too extreme to be 
compatible with well-mixed interstellar gas enriched by normal 
stellar populations. 
Shields (1996) noted that the BAL estimates are similar 
to the abundances measured in novae. He proposed that novae might 
dominate the enrichment of BAL regions if the nova rates are enhanced 
by white dwarfs gaining mass as they plunge through the QSO accretion 
disk. However, this unusual mechanism is not needed because there 
appear to be serious problems with the BAL column densities. 

\subsection{Problems with the BALs}

Absorption line optical depths and column densities 
are derived from the observed intensities, $I_{\nu}$, by the relation, 
$I_{\nu} = I_o\exp{(-\tau_{\nu})}$, where $I_o$ is 
the unabsorbed (continuum) intensity and $\tau_{\nu}$ is the 
optical depth at frequency $\nu$ (Junkkarinen \etal 1983, 
Grillmair \& Turnshek 1987, Korista \etal 1992, 
Savage \& Sembach 1991, Jenkins 1996, Arav \etal 1999). 
Partial coverage of the background light source(s) can 
fill in the troughs and thus lead to underestimated optical 
depths and column densities. 
A critical difference between BALs and NALs is in our ability 
to resolve adjacent multiplet lines and thereby directly measure 
any partial coverage effects (see Hamann \etal 1997b and HF99). 
In general, this type of analysis is possible for only the 
NALs, where resolved doublets like CIV~\lam\lam 1548,1550 or 
SiIV~\lam\lam 1394,1403 are easily measurable. The blending 
of these features in BALs has generally made the partial-coverage 
analysis impossible. 

Hamann (1998) used explicit calculations of the line optical 
depths to show that BAL spectra are consistent with much 
lower (solar) metallicities and solar relative abundances 
if at least the stronger transitions (such as CIV~\lam 1549) 
are more optically thick than they appear. In particular, 
those calculations suggest that the anomalously large ratio 
of PV~\lam 1121/CIV~\lam 1549 absorption lines results from 
severe saturation in CIV rather than extreme P/C abundances. 
The line saturation is disguised in the 
observed (modest) BAL troughs by partial line-of-sight 
coverage of the background light source(s). 

Arav \etal (1999) 
provided direct evidence for partial coverage in one 
BAL system based on some far-UV lines (whose true optical 
depth ratios are fixed by atomic physics). Similar evidence 
has come from a few BALs or BAL components where the profiles 
are narrow enough to resolve multiplet transitions that are 
close in wavelength (Telfer \etal 1998, Wampler \etal 1995). 
More indirect evidence for partial coverage in BALs 
has come from spectropolarimetry (Cohen \etal 1995, 
Goodrich \& Miller 1995, Hines \& Wills 1995). It is 
also interesting that the only known NAL system 
with PV absorption (Barlow \etal 1999) has resolved 
doublet ratios in CIV, SiIV and NV that require severe 
line saturation and partial line-of-sight coverage.  

It is therefore likely that the column densities derived 
so far from BALs are generally underestimated and the 
abundances are generally incorrect. Future studies 
that involve relatively narrow BALs (Telfer \etal 1998) 
or the plethora 
of hard-to-measure far-UV lines (Arav \etal 1999) might 
yet allow the needed partial-coverage analysis and thus 
provide reliable abundance estimates. 

\section{Metallicities and the Baldwin Effect}

The ``Baldwin Effect'' is an empirical 
trend for decreasing line equivalent widths with increasing 
luminosity (Baldwin 1977, Kinney \etal 1990, Osmer \& 
Shields this volume). Calculations by HF93a showed that most 
of the line equivalent widths will decline naturally with 
increasing metallicities (because of the declining gas temperatures, 
see also HF99). 
If $Z$ is correlated with the luminosities of QSOs, as suggested 
by the emission line data (\S4.2), then metallicity difference 
would at least contribute to the Baldwin Effect. 
An important test of the influence of metallicity 
will come from the nitrogen lines, because they should run 
counter to the trend and get relatively stronger with $Z$ 
(and $L$) due to the selective N enhancement (\S2.4 and 
\S3). Recent observations support this prediction, 
showing that while CIV and other lines decline with $L$ 
-- the usual Baldwin effect -- NV does not (Osmer \etal 
1994; Laor \etal 1995, Korista \etal 1998). See Korista 
\etal (this volume) for further discussion. 

\section{Summary and Future Prospects}

In spite of the null results from BALs, there is a growing 
consensus from the BELs and intrinsic NALs for typically 
solar or higher metallicities in high-redshift QSOs. These 
results support models of galaxy evolution wherein vigorous 
star formation in galactic nuclei (or dense proto-galactic 
condensations) produces $Z\ga$~\Zsun\ 
in the gas at redshifts $z\ga 4$. It is interesting 
to note that the high metal abundances 
in dense quasar environments sharply contrast with  
the metallicities measured elsewhere at high redshifts. 
For example, the damped-\Lya\ absorbers in QSO spectra, which 
apparently probe lines of sight through intervening disk 
galaxies (Prochaska \& Wolfe 1998 and refs. therein), 
have mean (gas-phase) metallicities of order 0.05~\Zsun\ 
at $z\sim 2$--3 (Lu \etal 1996, Pettini \etal 1997, 
Lu, Sargent \& Barlow 1998). The \Lya\ forest systems, 
which presumably probe more extended and tenuous inter-galactic 
structures (Rauch 1998), typically have metalicities $<$0.01~\Zsun\ 
at high redshifts (Rauch, Haehnelt \& Steinmetz 1997, 
Songalia \& Cowie 1996). 
The much higher metal abundances near QSOs are consistent 
with the rapid and more extensive evolution 
expected in dense environments (Gnedin \& Ostriker 1997). 

These are exciting times for quasar abundance work. 
The results so far have only scratched the surface of what 
is possible. The new generations of large ground-based and 
space-based telescopes are or will soon make it possible to 
greatly extend the results discussed above. 
In particular, we will be able to 1) 
measure a wider variety of both emission and absorption 
diagnostics and 2) compare the derived abundances in large QSO 
samples that span a wide range of redshifts and luminosities. 
The new data will thereby 
test further the reliability of each diagnostic, search 
more definitively for trends with redshift or luminosity, 
examine the range of QSO abundances at any given $z$ or $L$, 
and make better measurements of specific evolution probes like 
the Fe/$\alpha$ clock. It will be particularly interesting 
to compare emission and absorption diagnostics in the same 
objects. For example, one goal should be to test the BEL result 
for high Fe/Mg abundances by observing FeII/MgII 
or perhaps FeII/SiII in NAL systems. 

Inevitably, these studies will also improve our general understanding 
of observational trends between QSO luminosities and the BEL 
spectra. Abundance variations might be an important ingredient 
in calibrating the Baldwin Effect for tests 
of cosmology (\S6, also Korista \etal this volume).

\acknowledgments
I am greatful to my close collaborators, T. Barlow, F. Chaffee, 
G. Ferland, C. Foltz, V. Junkkarinen, K. Korista and J. Shields, 
for their contributions to this work. I also acknowledge support 
from the Space Telescope Guest Observer program and from NASA through 
grant NAG 5-3234.

\end{document}